\documentclass[prl,twocolumn,groupedaddress]{revtex4}

\usepackage[dvips]{graphicx}

\begin{document}

\newcommand{\av}[1]{\left\langle#1\right\rangle}
\newcommand{\£}{\pounds}
\newcommand{\E}{\mathrm{E}}
\newcommand{\Var}{\mathrm{Var}}
\newcommand{\Cov}{\mathrm{Cov}}


\title{Detecting a Currency's Dominance or Dependence \\
using Foreign
Exchange  Network Trees}
\author{Mark McDonald$^{1}$, Omer Suleman$^2$, Stacy Williams$^3$, Sam Howison$^1$ 
and Neil F. Johnson$^2$}
\address{$^1$Mathematics Department, Oxford University, Oxford, OX1 2EL, U.K. \\
$^2$Physics Department, Oxford University, Oxford OX1 3PU, U.K.  \\
$^3$FX Research and Trading Group, HSBC Bank, 8 Canada Square, London E14 5HQ, 
U.K.}

\date{\today}

\begin{abstract} In a system containing a large number of interacting stochastic processes, there will typically be many non-zero correlation coefficients. This makes it difficult to either visualize the system's inter-dependencies, or identify its dominant elements. Such a situation arises in Foreign Exchange (FX) which is the world's biggest market. Here we develop a network analysis of these correlations using  Minimum Spanning Trees (MSTs). We show that not only do the MSTs provide a meaningful representation of the global FX dynamics, but they also enable one to determine momentarily dominant and dependent currencies. We find that information about a country's geographical ties emerges from the raw exchange-rate data. Most importantly from a trading perspective, we discuss how to infer which currencies are `in play' during a particular period of time.
\end{abstract}

\maketitle

\section{I. Introduction} 
There is enormous interest in the properties of complex networks \cite{nets1,nets2,DM2003}. There has been an explosion of papers within the physics literature analyzing the structural properties of biological, technological and social networks; the main results of which are summarized in \cite{DM2003}. Such networks or `graphs', contain $n$ nodes or `vertices' $\{i\}$ connected by $M$ connections or `edges'. In the case of physical connections, such as wires or roads, it is relatively easy to assign a binary digit (i.e. 1 or 0) to the edge between any two nodes $i$ and $j$  according to whether the corresponding physical connection exists or not. However, for social networks such as friendship networks \cite{DM2003}, and biological networks such as reaction pathways \cite{DM2003}, the identification of network connections is less clear. In fact it is extremely difficult to assign any particular edge as being a definite zero or one -- instead, all edges will typically carry a
weighting value $\rho_{ij}$ which is analog rather than binary, and which is in general neither
equal to zero nor to one. The analysis of such weighted networks is in its infancy, in particular with respect to their functional properties and dynamical evolution \cite{econo}.  The main difficulty is that the resulting network is fully-connected with $M=n(n-1)$ connections between all $n$ nodes. In symmetric situations where $\rho_{ij}\equiv
\rho_{ji}$, this reduces to $M=n(n-1)/2$ connections, but is still large for any reasonable $n$. 

An interesting example of such a fully-connected weighted network is provided by the set of correlation coefficients between $n$ stochastic variables. Each node $i$ corresponds to the stochastic variable $x_i(t)$ where $i=1,2,\dots,n$, and each of the $n(n-1)/2$ connections between pairs of nodes carries a weight given by the value of the correlation coefficient $\rho_{ij}$ (see definition below). For any reasonable number of nodes the number of connections is very large (e.g.  for $n=110$, $n(n-1)/2=5995$) and hence it is extremely difficult to deduce which correlations are most important for controlling the overall dynamics of the system. Indeed, it would be highly desirable to have a simple method for deducing whether certain nodes, and hence a given subset of these stochastic processes, are actually `controlling' the correlation structure \cite{note9}. In the context of financial trading, such nodal control would support the popular notion among traders that certain currencies can be `in play'
over a given time period. Clearly such information could have important practical consequences in terms of understanding the overall dynamics of the highly-connected FX market. It could also have practical applications in other areas where $n$ inter-correlated stochastic process are operating in parallel. 

With this motivation, we present here an analysis of the correlation network in an important real-world system, namely the financial currency (i.e. FX) markets. Although the empirical analysis presented is obtained specifically for this financial system, the analysis we provide has more general relevance to any system involving $n$ stochastic variables and their $n(n-1)/2$ correlation coefficients. There is no doubt that currency markets are extremely important \cite{PEV2003} -- indeed, the recent fall in the value of the dollar against other
major currencies is quite mysterious, and has attracted numerous economic `explanations' to reason away its dramatic decline. The currency markets, which represent the largest market in the world, have daily transactions totalling trillions of dollars, exceeding the yearly GDP (Gross Domestic Product) of most countries. 

The technical approach which we adopt, is motivated by recent research within the physics community by Mantegna and others \cite{MRN1999,MS2000,JPO2002,OCKK2003b} and concerns the construction and analysis of Minimum Spanning Trees (MST), which contain only $n-1$ connections. Mantegna and co-workers focused mainly on equities -- by contrast, we consider the case of FX markets and focus on  what the time-dependent properties of the MST can tell us about the FX market's evolution. In particular, we investigate the stability and time-dependence of the resulting MST, and introduce a methodology for inferring which currencies are `in play' by analyzing the clustering and leadership structure within the MST network. 

The application of MST analysis to financial stock (i.e. equities) was introduced by the physicist Rosario Mantegna \cite{MRN1999}. The MST gives a `snapshot' of such a system; however, it is the temporal evolution of such systems, and hence the evolution of the MSTs themselves, which motivates our research. In a series of papers \cite{OCKK2003b,OCKK2003a,OCKK2003c}, Onnela et al. extended Mantegna's work to investigate how such trees evolve over time in equity markets. Here we follow a similar approach for FX markets.  One area of particular interest in FX trading -- but which is of interest for correlated systems in general -- is to identify which (if any) of the currencies  are `in play' during a given period of time. More precisely, we are interested in understanding whether particular currencies appear to be assuming a dominant or dependent role within the network, and how this changes over time. Since exchange rates are always quoted in terms of the price of one currency compared to another, this is a highly non-trivial task. For example, is an increase of the value of the euro versus the dollar primarily because of an increase in the intrinsic value of euro, or a decrease in the intrinsic value of the dollar, or both? We analyze FX correlation networks in an attempt to address such questions. We believe that our findings, while directly relevant to FX markets, could also be relevant to other complex systems containing $n$ stochastic processes whose interactions evolve over time.

\section{II. Minimum Spanning Tree (MST)} 
Given a correlation matrix (e.g. of financial returns) a connected graph can be constructed by means of a transformation between correlations and suitably defined distances \cite{MS2000}. This transformation assigns smaller distances to larger correlations \cite{MS2000}. The MST, which only contains $n-1$ connections, can then be constructed from the resulting hierarchical graph \cite{RTV1986,MS2000}. Consider $n$ different time-series, $x_i$  where $i\in \{1,2,...n\}$, with each time-series $x_i$ containing $N$ elements (i.e. $N$ timesteps). The corresponding $n \times n$ correlation matrix $C$ can easily be constructed, and has elements $C_{ij}\equiv\rho_{ij}$ where 
\begin{eqnarray} \rho_{ij} & = & \frac{ \langle x_i x_j \rangle - \langle x_i \rangle \langle x_j \rangle}{\sigma_i \sigma_j}
\end{eqnarray} 
where $\langle$...$\rangle$ indicates a time-average over the $N$ datapoints for each $x_i$, and
$\sigma_i$ is the sample standard deviation of the time-series $x_i$. From the form of $\rho_{ij}$ it is obvious that $C$ is a symmetric matrix. In addition, 
\begin{eqnarray} 
\rho_{ii} = \frac{\langle x_i^2 \rangle - \langle x_i \rangle ^2}
{\sigma_i^2} & \equiv & 1,  \ \ \forall\  i
\end{eqnarray} 
hence all the diagonal elements are identically 1. Therefore $C$ has $n(n-1)/2$ independent elements. Since the number of relevant correlation coefficients increases like $n^2$, even a relatively small number of time-series can yield a correlation matrix which contains an enormous amount of information -- arguably `too much' information for practical
purposes. By comparison, the MST provides a skeletal structure with only $n-1$ links, and hence attempts to strip the system's complexity down to its bare essentials. As shown by Mantegna, the practical justification for using the MST lies in its ability to provide economically meaningful information \cite{MRN1999,MS2000}. Since the MST contains only a subset of the information from the correlation matrix, it cannot tell us anything which we could not (in principle) obtain by analyzing the matrix $C$ itself. However, as with all statistical tools, the hope is that it can provide an insight into the system's overall behavior which would not be
so readily obtained from the (large) correlation matrix itself.

To construct the MST, we first need to convert the correlation matrix $C$ into a `distance' matrix $D$. Following Refs. \cite{MRN1999,MS2000}, we use the non-linear mapping:
\begin{eqnarray} 
d_{ij}(\rho_{ij}) &=& \sqrt{2(1-\rho_{ij})}
\end{eqnarray} 
to get the elements $d_{ij}$ of $D$ \cite{note1}. Since $-1\leq \rho_{ij} \leq 1$, we have
$0\leq d_{ij} \leq 2$. In particular:
\begin{eqnarray*}
\rho_{ij} = -1 & \longmapsto & d_{ij} = 2 \\
\rho_{ij} = 0  & \longmapsto & d_{ij} = \sqrt{2}\\
\rho_{ij} = \frac{1}{2} & \longmapsto & d_{ij} = 1\\
\rho_{ij} = 1  & \longmapsto & d_{ij} = 0
\end{eqnarray*}
This distance matrix $D$ can be thought of as representing a fully connected graph with edge weights $d_{ij}$. In the terminology of graph theory, a `forest' is a graph where there are no cycles \cite{BBo1979} while a `tree' is a connected forest. Thus a tree containing $n$ nodes must contain precisely $n-1$ edges \cite{DM2003,BBo1979}. The minimum spanning tree $\bf{T}$ of a graph is the tree containing every node, such that the sum $\sum_{d_{ij}\in \bf{T}}{d_{ij}}$ is a minimum. There are two methods for constructing the MST --- Kruskal's algorithm and Prim's algorithm \cite{JPO2002}. We used Kruskal's algorithm, details of which are given in \cite{CLR1990}.

\section{III. Data}
\subsection{1. Raw Data} 
The empirical currency data that we investigated are hourly, historical price-postings from  HSBC Bank's database for nine currency pairs together with the price of gold from 01/04/1993 to 12/30/1994 \cite{note8}.  Gold is included in the study because there are similarities in the way that it is traded, and in some respects it resembles a very volatile currency. The currency pairs under investigation are AUD/USD, GBP/USD, USD/CAD,	USD/CHF,	USD/JPY, GOLD/USD, USD/DEM, USD/NOK, USD/NZD, USD/SEK \cite{note2}. In the terminology used in FX markets \cite{note2}, USD/CAD is counter-intuitively the number of Canadian dollars (CAD) that can be purchased with one US dollar (USD). We must define precisely what we mean by hourly data, as prices are posted for different currency pairs at different times. We do not want to use average prices since we want the prices we are investigating to be prices at which we could have executed trades. Hence for hourly data, we use the last posted price within a given hour to represent the (hourly) price for the following hour. 

We emphasize that the $n$ stochastic variables which we will analyze correspond to currency {\em exchange} rates and hence measure the {\em relative} values of any two currencies. It is effectively meaningless to ask the {\em absolute} value of a given currency, since this can only ever be measured with respect to some other financial good. Thus each currency pair corresponds to a node in our network. We are concerned with the correlations between these currency exchange rates, each of which corresponds to an edge between two nodes. A given node does {\em not} correspond to a single currency.

\subsection{2. Data Filtering} 
As with all real-world systems, the issue of what constitutes correct data is complicated. In particular, there are some subtle data-filtering (or so-called `data-cleaning') issues which need to be addressed. Such data problems are, by contrast to the physical sciences, a reality in most disciplines which deal with human timescales and activity. However, rather than being a reason not to work in such areas, we feel that physicists can offer insight into the extent to which such missing data is likely to be important for understanding a given system's overall dynamics.  In our specific case, we are interested in calculating both the instantaneous and lagged correlations between exchange-rate returns. Hence it is neccessary to ensure that (a)
each time series has an equal number of posted prices; (b) the $n$'th posting for each currency pair corresponds, to as good an approximation as possible, to the price posted at the same timestep $t_n$ for all $n \in \{1, ... , N\}$. For some of the hourly timesteps, some currency pairs have missing data. The best way to deal with this is open to interpretation. Is the data missing simply because there has been no price change during that hour, or was there a fault in the data-recording system? Looking at the data, many of the missing points do seem to occur at times when one might expect the market to be illiquid. However, sometimes there are many consecutive missing data points --- even an entire day. This obviously reflects a fault in
the data recording system. To deal with such missing data we adopted the following protocol.  The FX market is at its most liquid between the hours of 08:00 and 16:00 GMT \cite{JJPS2004}. In an effort to eradicate the effect of `zero returns' due to a lack of liquidity in the market -- as opposed to the price genuinely not moving in consecutive trades -- we only used data from between these hours \cite{note5}. Then, if the missing data were for fewer than three consecutive hours, the missing prices were taken to be the value of the last quoted price. If the missing data were for three or more consecutive hours, then the data for those hours were omitted from the analysis. Since we must also ensure completeness of the data at each point, it is then necessary that the data for those hours are omitted from {\em all} currency pairs under investigation \cite{note6}. We believe that this procedure provides a sensible compromise between the conflicting demands of incorporating all relevant data, and yet avoiding the inclusion of spurious zero-returns which could significantly skew the data.

\subsection{3. Stationarity} Once we have cleaned up the exchange-rate data for each currency pair, whose associated price we will henceforth label as $P_i(t)$, we  turn this value into a financial return
\begin{eqnarray} 
r_i(t_n) & = & \ln \left( \frac{P_i(t_{n+1})}{P_i(t_n)}\right).
\end{eqnarray}

\begin{figure}[ht]
\includegraphics[width=.47\textwidth]{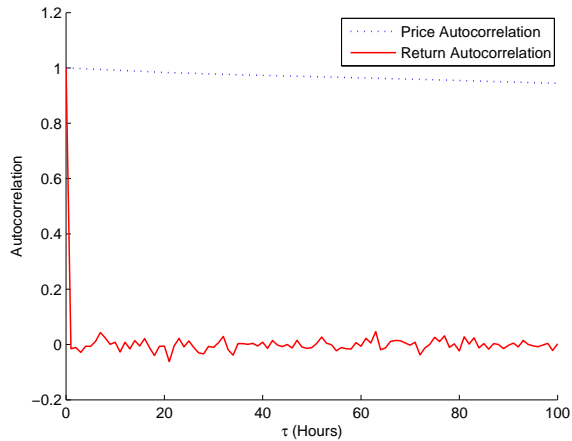}
\caption{Autocorrelation of the price and return for the currency pair AUD/USD (i.e.
the Australian dollar versus the US dollar).}
\label{AutoCorrel}
\end{figure}

Since we will be calculating correlations, we need to be confident that our return distributions are stationary. A useful probe of stationarity is to calculate the autocorrelation for each time-series. A stationary time-series will have an autocorrelation function which rapidly decays to zero \cite{PR1998}, whereas a non-stationary time-series will have an autocorrelation function which decays to zero very slowly (if at all). The autocorrelation is defined as
\begin{eqnarray}
\rho_{i}(\tau) &=& \frac{\av{x_i(t+\tau)x_i(t)} -
\av{x_i(t+\tau)}\av{x_i(t)}}{\sigma_{i,\tau}\sigma_i}
\end{eqnarray} where $\av{...}$ indicates a time-average over the $N-\tau$ elements and $\sigma_{i,\tau}$, $\sigma_i$ are the sample standard deviations of the time-series $x_i(t+\tau)$ and $x_i(t)$, respectively. To illustrate this point, Figure \ref{AutoCorrel} shows the autocorrelation of both the price and return for the currency pair AUD/USD over the two year period of interest. It can be clearly seen that this confirms the assertion that the returns are stationary whilst the prices themselves are not. Thus we can, with confidence, calculate correlations between different currency-pair returns.

\subsection{4. Foreign Exchange Data} In addition to the problems outlined above which are common to the analyses of all such real-world data, there are further issues that are specific to FX data and which make the study of FX and equities fundamentally different. When producing the MST for the returns of the stock which make up the FTSE100 index, one calculates the returns
from the values of the price of the stock \emph {in the same currency} --- specifically, UK pounds (GBP). With FX data, however, we are considering exchange rates between currency pairs. Thus should we consider GBP/USD or USD/GBP? And does it indeed make a difference which one we use? Since the correlation is constructed to be normalized and \emph{dimensionless}, one might be tempted to think that it does not matter since the value of the correlation will be the same and only the sign will be different. However, it is important when constructing the MST since there is an asymmetry between how positive and negative correlations are represented as distances. In particular, the MST picks out the smallest distances --- i.e. the highest correlation. A large negative correlation gives rise to a large distance between nodes. Thus a connection between two nodes will be missing from the tree even though it would be included if the other currency in the pair were used as the base currency.

Consider the following example. There is a large \emph{negative} correlation between the returns of the two currency pairs GBP/USD and USD/CHF \cite{note3}. Conversely, if we put them both with USD as the base currency, we get a large \emph{positive} correlation between USD/GBP and USD/CHF. Thus our choice will give rise to a fundamentally different tree structure.
For this reason, we perform the analysis for all possible currency-pairs against each other. Since we are analysing ten currency pairs, this gives us eleven separate currencies and hence 110 possible currency pairs (and hence $n=110$ nodes). However, there are constraints on these timeseries and hence an intrinsic structure is imposed on the tree by the relationships between the timeseries. This is commonly known as the `triangle effect'. Consider the three exchange-rates USD/CHF, GBP/USD and GBP/CHF. The $n$th element of the timeseries for GBP/CHF is simply the product of the $n$th elements of USD/CHF and GBP/USD. This simple relationship between the timeseries gives rise to some relationships between the correlations. More generally, with three time-series $P_1(t)$,$P_2(t)$,$P_3(t)$ such that $P_3(t) = P_1(t)P_2(t)$, there exist relationships between the correlations and variances of the returns. If we define
the returns $r_i$ such that $r_i = \ln P_i$ for all $i$, then we have:
\begin{eqnarray} 
r_3 = r_1 + r_2\ .
\end{eqnarray} Thus
\begin{eqnarray}
\Var(r_3) &=& \Var (r_1+r_2)\\
 &=& \E((r_1+r_2)^2) - (\E(r_1+r_2))^2
\end{eqnarray} 
For currency pairs, it is valid to assume that the expected value of the return is zero \cite{note4}. Hence this expression simplifies to
\begin{eqnarray}
\sigma_3^2 &=& \E(r_1^2+r_2^2+2r_1r_2)\\
 &=& \sigma_1^2 + \sigma_2^2 + 2\Cov(r_1,r_2)\\
 &=& \sigma_1^2 + \sigma_2^2 + 2\sigma_1\sigma_2\rho_{12}
\label{good}
\end{eqnarray} where $\sigma_1$, $\sigma_2$, $\sigma_3$ are the variances of the returns $r_1(t)$, $r_2(t)$, $r_3(t)$ while $\rho_{12}$ is the correlation between the returns $r_1(t)$ and $r_2(t)$. Finally we obtain
\begin{eqnarray}
\rho_{12} &=& \frac{\sigma_3^2 - (\sigma_1^2+\sigma_2^2)}{2\sigma_1\sigma_2}
\label{Triangle}
\end{eqnarray}
Hence there is a structure forced upon the market by the triangle effect. This is not a problem since all the cross-rates we include in the tree \emph{do} exist and the correlations calculated are the true correlations between the returns. Even though the values of these correlations have some relationships between them, they should be included in the tree since it is precisely this market structure that we are attempting to identify. We do, however, need to confirm that this structure which is being imposed on the market is not dominating our results. 

\subsection{5. Checking Data Validity} 
\begin{figure}[h]
\includegraphics[width=.47\textwidth]{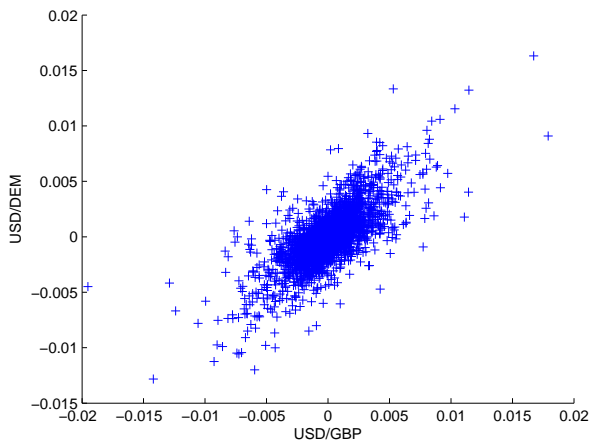}
\caption{Scatter plot of returns for USD/DEM (i.e. the US dollar versus the German
mark) against USD/GBP (i.e. the US dollar versus the UK pound) }
\label{scatter}
\end{figure}

We performed a simple check on the data by calculating the minimum and maximum return for each currency pair. For example, if the rate for USD/JPY was entered as 1.738 instead of 173.8 then this would give rise to returns of approximately $\pm 1$. As a result of this check, we could confirm that there were no such errors in our dataset. We then drew scatter plots of the returns against time, plus histograms of the return distribution in order to check  that there were no unusual points on the graph. The next check that we performed is slightly more subtle. The correlation between two variables is related to the gradient of the regression line between the variables \cite{PR1998}. However, this gradient is very susceptible to outliers so we need to ensure that our data does not contain such outlying points. To check this, one can plot scatter-graphs of returns from different currency pairs against each other. Since we have 110 currency pairs, there are 5995 possible scatter-graphs (one for each of the $n(n-1)/2$ independent correlation coefficients). Obviously it is unfeasible to plot all of these. However, as we have seen, all the cross-rates are formed from the USD rates. Therefore we can just plot the 45 scatter-graphs of USD/A against USD/B where A $\neq$ B. Figure \ref{scatter} shows this scatter plot for USD/DEM vs USD/GBP. It can be seen that there are no such outlying points. This figure is typical of all the 45 possible plots.

\section{IV. Directed Trees} 
\begin{figure}[ht]
\includegraphics[width=.47\textwidth]{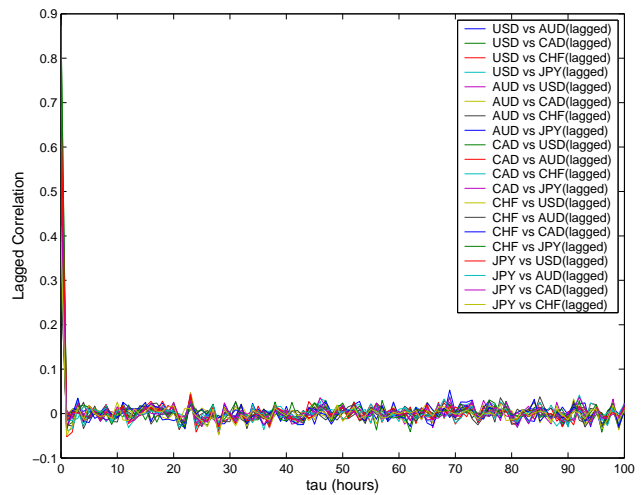}
\caption{Lagged correlation between different currency pairs when GBP is the base
currency. As explained in the text, AUD vs USD (lagged) refers to the lagged
correlation between GBP/AUD (at time $t+\tau$) and GBP/USD (at time $t$)}
\label{LagGBP}
\end{figure}

In \cite{KKK2003}, the Minimum Spanning Tree approach was generalized by considering a directed graph. Lagged correlations were investigated in an attempt to determine whether the movement of one stock price `preceded' the movement in another stock price. We now investigate whether this approach yields useful results here. First we should define what we mean by lagged correlation. If we have two time-series, $x_i(t)$ and $x_j(t)$ where both time-series contain $N$ elements, the $\tau$-lagged correlation is given by
\begin{eqnarray} 
\rho_{ij}(\tau) &=& \frac{\langle x_i(t+\tau)x_j(t) \rangle - \langle x_i(t+\tau)
\rangle \langle x_j(t) \rangle}{\sigma_{i,\tau} \sigma_j}
\end{eqnarray} where $\langle$...$\rangle$ indicates a time-average over the $N-\tau$ elements and $\sigma_{i,\tau}$, $\sigma_j$ are the sample standard deviations of the time-series $x_i(t+\tau)$ and $x_j(t)$, respectively. Note that autocorrelation is simply the special case of this where $i=j$. Armed with this definition, we can now look at our data to see whether there are any significant lagged correlations between returns of different currency pairs. Figure \ref{LagGBP} shows the lagged correlation between the returns of each pair of currencies when the prices of those currencies are given with GBP as the base currency. In the figure, AUD vs USD (lagged) refers to the lagged correlation between GBP/AUD (at time $t+\tau$) and GBP/USD (at time $t$). The results in this figure are representative of the results from all currency pairs included in our study.

Figure \ref{LagGBP} clearly shows that the approach considered in \cite{KKK2003} will not yield anything useful here for FX. If such lagged correlations do exist between currency pairs, they occur over a timescale smaller than one hour. In other words, the FX market is very efficient. This should not come as a surprise --- the FX market is approximately 200 times as liquid as the equities market \cite{PEV2003}.

\section{V. The Currency Tree} 
\begin{figure}[ht]
\includegraphics[width=.47\textwidth]{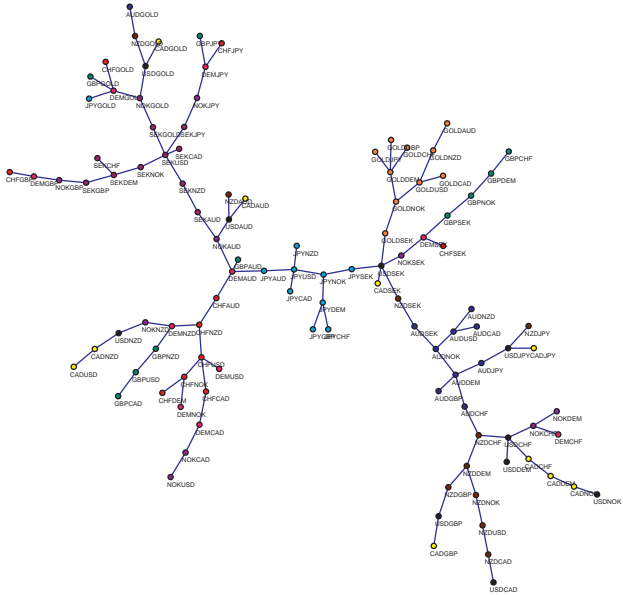}
\caption{The Minimum Spanning Tree representing the correlations between all hourly
cross-rate returns from the years 1993 and 1994. Created using Pajek software.}
\label{All9394}
\end{figure}

Creating all the possible cross-rates from the 11 currency pairs gives rise to a total of $n=110$ different time-series. It is here that the approach of constructing the MST comes into its own, since 110 different currencies yields an enormous correlation matrix containing 5995 separate elements. This is far too much information to allow any practical analysis by eye. However, as can be seen from Figure \ref{All9394}, the hourly FX tree is quite easy to look at. Rather than a mass of numbers,  we now have a graphical representation of the complex system in
which the structure of the system is visible.

\begin{figure}[ht]
\includegraphics[width=.47\textwidth]{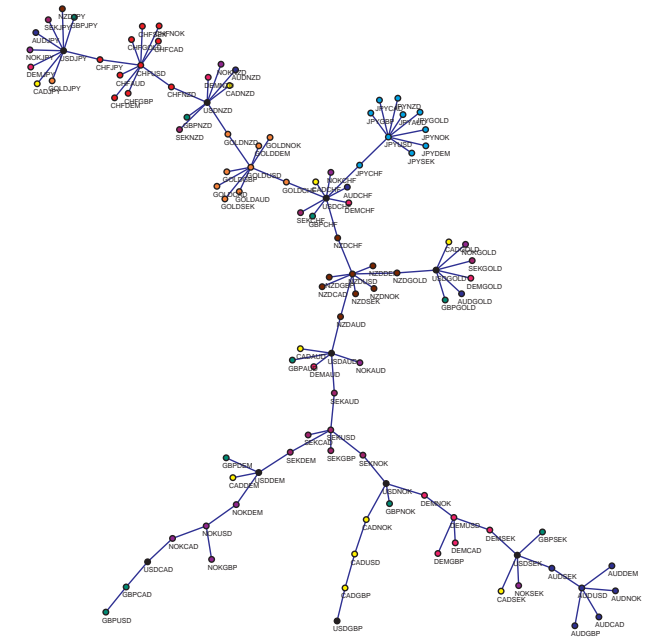}
\caption{The Minimum Spanning Tree formed from randomized data for the USD prices.
This shows only the structure imposed on the tree by the triangle effect}
\label{jumbled}
\end{figure}

Before analysing the tree in detail, it is instructive to consider first what effect the constraints of Eq. (12)  (the `triangle effect') will have on the shape of the tree. Figure \ref{jumbled} illustrates this. The data used in this figure is the same data as in Fig. \ref{All9394}, however the price-series for the currencies in USD were randomized before the cross-rates were formed. This process gives prices for the various currencies in USD which are random, and will hence have negligible correlation between their returns. Thus Fig. \ref{jumbled} shows the structure forced on the tree by the triangle effect.

\begin{figure}[ht]
\includegraphics[width=.47\textwidth]{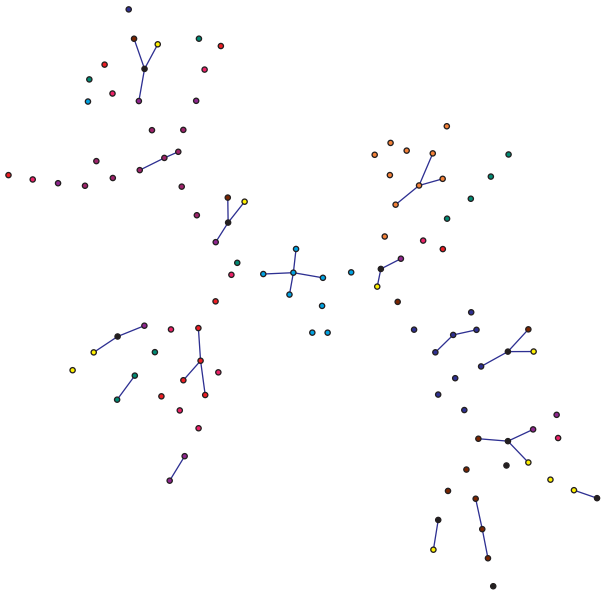}
\caption{Figure \ref{All9394} redrawn with only the links which are present in both
Figs. \ref{All9394} and \ref{jumbled}}
\label{Intersection}
\end{figure}

This tree resulting from randomizing data as described above, is actually very different in character from the true tree in Figure \ref{All9394}. At first glance it might appear that some aspects are similar --- currencies show some clustering in both cases. However, in the tree of real cross-rates there are currency-clusters forming about any node, whereas in Figure
\ref{jumbled} there are only clusters centred on the USD node. This is not surprising: after all, what do the `CHF/everything' rates all have in common in the case of random prices other than the CHF/USD rate?  The best way to interpret Fig. \ref{jumbled} is that we have a tree of USD nodes (which are spaced out since their returns are poorly correlated) and around these nodes we have clusters of other nodes which have the same base currency, and which are effectively the information from the USD node plus noise. This exercise shows us that the MST results are not dominated by the triangle effect. In an effort to show this in a more quantitative way, we investigate the proportion of links that are present in both trees. Less than one third of the edges in Fig. \ref{All9394} are present in Fig. \ref{jumbled}.  In Fig. \ref{Intersection} the same nodes are shown, but we have only included the links which are present in both Figs. \ref{All9394} and \ref{jumbled}.

\begin{figure}[ht]
\includegraphics[width=.47\textwidth]{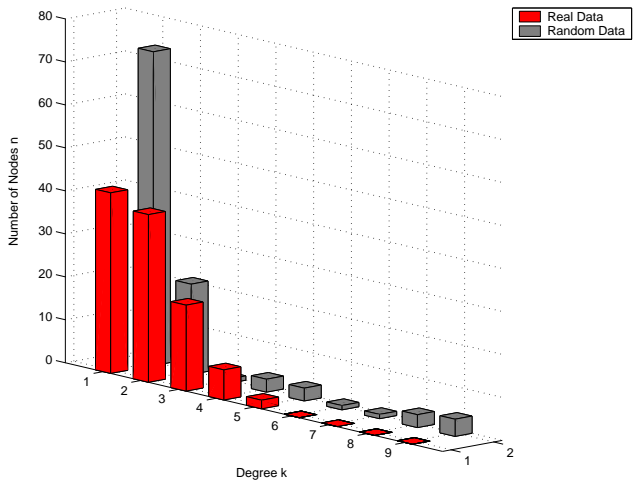}
\caption{Comparison of the degree distributions for the trees shown in Figure
\ref{All9394} (Real Data) and Figure \ref{jumbled} (Randomized Data)}
\label{degreedist}
\end{figure}

Another more quantitative comparison is to compare the degree distribution of the tree from the random price series with that of the tree from real price data. This is detailed in Table 1 and shown graphically in Fig. \ref{degreedist}. Again, this further highlights the differences between the two trees.
\begin{table}[h]
\begin{center}
\begin{tabular}{c|c|c}
\hline Degree(k) & Figure \ref{All9394} & Figure \ref{jumbled}\\
\hline 1 & 42 & 73 \\ 2 & 39 & 21 \\ 3 & 20 & 1 \\ 4 &  7 & 3 \\ 5 &  2 & 3 \\ 6 &  0
& 1 \\ 7 &  0 & 1 \\ 8 &  0 & 3 \\ 9 &  0 & 4 \\
\hline
\end{tabular}
\end{center}
\caption{Comparing the degree distibutions for Fig. \ref{All9394} and Fig.
\ref{jumbled} }
\end{table}

\begin{figure}[ht]
\includegraphics[width=.47\textwidth]{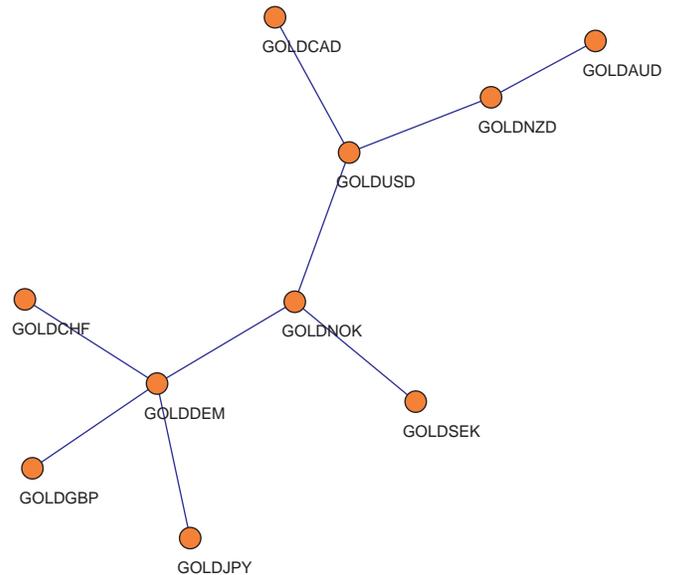}
\caption{The cluster of GOLD exchange-rates from Fig. \ref{All9394}}
\label{Gold}
\end{figure}

\begin{figure}[ht]
\includegraphics[width=.47\textwidth]{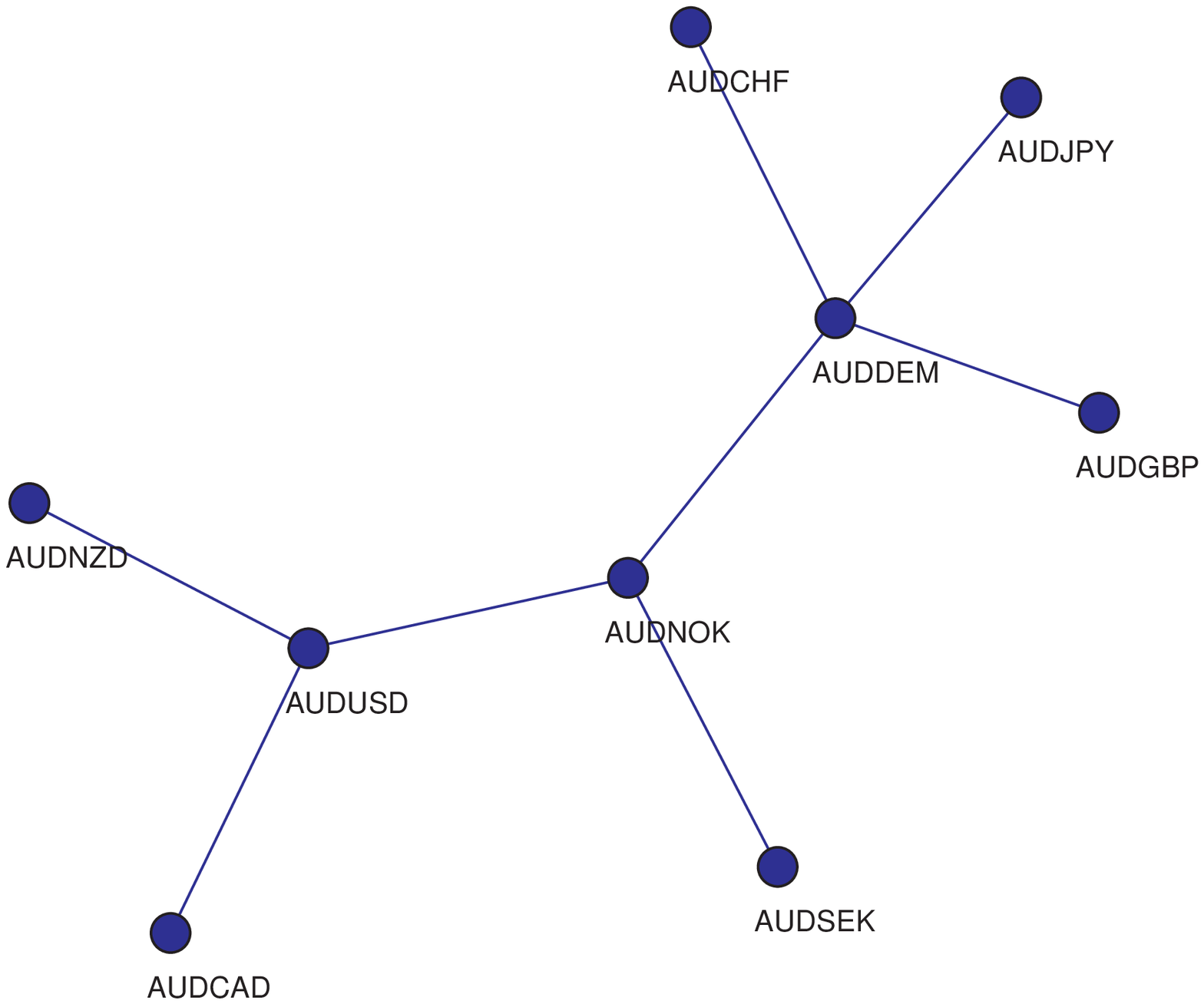}
\caption{The cluster of AUD exchange-rates from Fig. \ref{All9394}}
\label{AUD}
\end{figure}

It is encouraging that the cluster of Gold exchange rates links currencies in a sensible way. This cluster is re-drawn in Figure \ref{Gold}. It can be seen that the nodes in this cluster are grouped in an economically meaningful way: remarkably, there is a geographical linking of exchange-rates. The Australisian nodes, AUD and NZD, are linked, as are the American ones (USD and CAD). The Skandinavian currencies, SEK and NOK, are also linked. Finally, there is a European cluster of GBP, CHF and EUR. A similar effect is noticeable in the AUD cluster (shown again in Figure \ref{AUD}). This provides a useful check that our results are sensible. Indeed if such geographical clustering had not arisen, it would be a good indication that something was wrong with our methodology.

\section{VI. Stability and Temporal Evolution of the Currency Tree} 
\begin{figure}[ht]
\includegraphics[width=.47\textwidth]{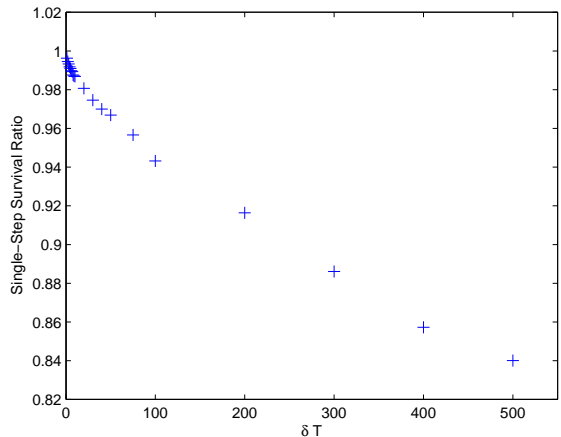}
\caption{Single Step survival ratio as a function of $\delta T$}
\label{SingleStep}
\end{figure}

We now investigate the single-step survival ratio of the edges (i.e. connections)
\begin{eqnarray}
\sigma_{\delta t} &=& \frac{\left| E_t\cap E_{t+\delta t} \right|}{\left| E
\right|}
\end{eqnarray} where $E_t$ and $E_{t+\delta t}$ represent the set of edges (i.e. connections) present at times $t$ and $t+\delta t$ respectively, in order to see how it depends on the value chosen for $\delta t$. This ratio must tend to one as $\delta t$ approaches 0 for our results
to be meaningful. The results are plotted in Fig. \ref{SingleStep} and it can be seen that it is indeed the case that this ratio tends to one as $\delta t$ approaches 0. Thus the topology of the MST is stable.

Next we investigate the time-dependence of the tree. Onnela \cite{JPO2002} defined the $k$ multi-step survival ratio to be:
\begin{eqnarray}
\sigma_{\delta t,k} &=& \frac{\left| E_t \cap E_{t+\delta t} \cap \ldots \cap
E_{t+k\delta t}  \right|}{\left| E \right|}
\label{min}
\end{eqnarray} Thus if a link disappears for only one of the trees in the time $t$ to $t+\delta t$ and then comes back, it is not counted in this survival ratio. This seems a possibly overly-restrictive definition which might underestimate the survival. We will therefore also consider the more generous definition
\begin{eqnarray}
\sigma_{\delta t,k} &=& \frac{\left| E_t \cap E_{t+k\delta t} \right|}{\left| E
\right|}\ \ .
\label{max}
\end{eqnarray} 
This quantity will, for large values of $k$, include cases where the links disappear and then come back several timesteps later. It therefore tends to overestimate the survival since a
reappearance after such a long gap is more likely to be caused by a changing structure than by a brief, insignificant fluctuation.

\begin{figure}[ht]
\includegraphics[width=.47\textwidth]{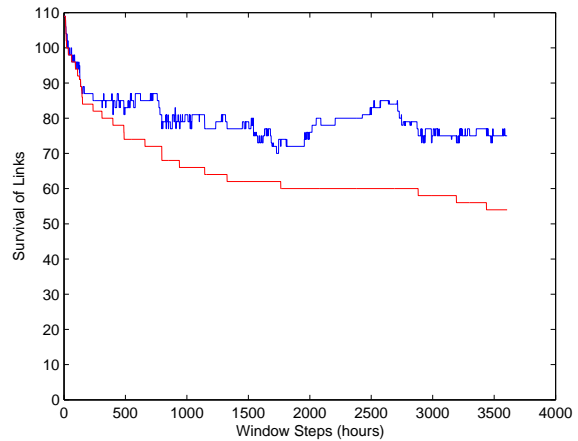}
\caption{Multi-step survival ratio of the FX tree's connections, as a function of
time. The graph shows the two definitions described in the text, which tend to
overestimate (blue) and underestimate (red) the survival effect.}
\label{MultiStep}
\end{figure}

Figure \ref{MultiStep} shows both definitions, and  uses a time-window of length $T=1000$ hours and a time-step $\delta t=1$ hour \cite{note7}. It can be seen from the figure that the two lines form a `corridor' for the multi-step survival ratio. This is because the over-restrictive definition of Eq. (\ref{min}) under-estimates the survival and the over-generous definition of Eq. (\ref{max}) over-estimates the result. It is particularly noteworthy that even with the over-restrictive definition of Eq. (\ref{min}), the survival of links after the end of two years is only just below fifty percent (i.e. 54/109). In other words, there are strong correlations existing between exchange-rate returns that are extremely long-lived.

\section{VII.  Interpreting the trees} 
We know from our analysis that clusters occur in the MST, and that these clusters change over time. Next we need to understand the significance of this in practice. We will approach this by analyzing two trees which are one calendar month apart.

\begin{figure}[ht]
\includegraphics[width=.47\textwidth]{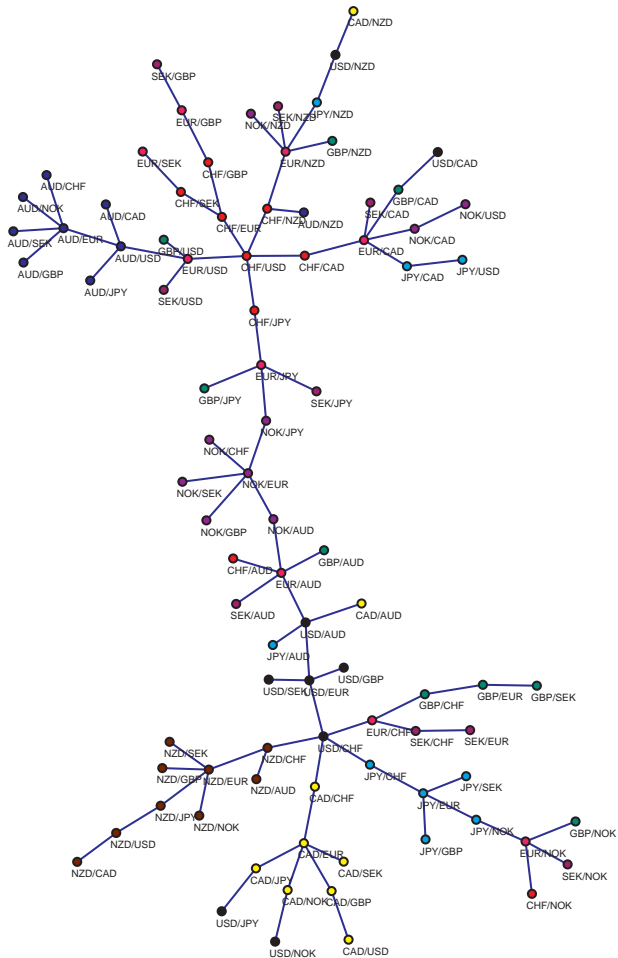}
\caption{Currency tree (MST) for a 2 week period in June 2004.}
\label{June}
\end{figure}

\begin{figure}[ht]
\includegraphics[width=.47\textwidth]{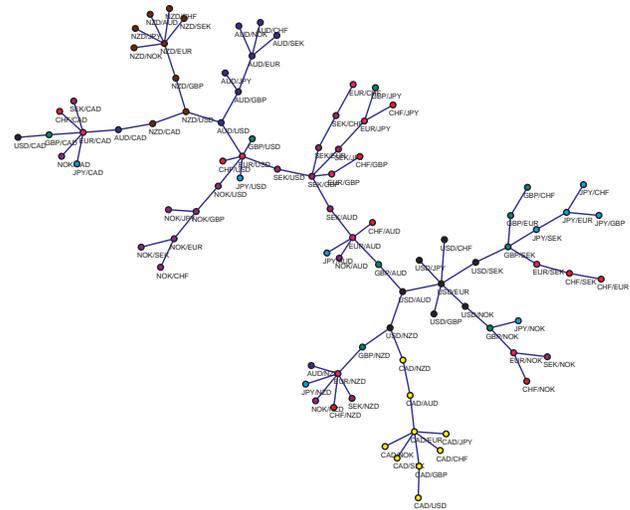}
\caption{Currency tree (MST) for a 2 week period in July 2004.}
\label{July}
\end{figure}

Figure \ref{June} shows an example of a currency tree from a period in June 2004. Despite the initial impression, the tree is actually very easy to interpret. It contains coloured points, each of which represents a particular currency pair. For the reasons explained earlier, currency pairs are quoted both ways round: USD/JPY appears with USD as the base currency, as is normal market convention, but so does JPY/USD. This gives all currencies the chance to stand out as a cluster, as will be seen shortly.
The currency-pair nodes are each colour coded, according to the labelled base currency. Broadly speaking, each node is linked to the node representing the currency-pair to which it is most closely correlated. The observation that certain currency-pairs cluster together means that they have been moving together consistently over the monitored period.

Figure \ref{June} shows a strong, yellow, CAD cluster near the bottom of the tree. Seven currency-pairs with CAD as the base currency are all connected, demonstrating that the Canadian dollar has been moving uniformly and systematically against a whole range of other currencies during that time. The CAD is `in play', to use the prevailing industry term. The same is true for the brown-coloured New Zealand dollar (NZD), which has also formed a cluster. Other clusters are also evident, including a red Swiss franc (CHF) cluster, which has formed near the top of the tree. In contrast, the Sterling currency-pairs are dispersed around the tree, indicating that there is little in common in their behaviour. In short, Sterling is 
not `in play'.

If the trees were static, this would be the end of the story. However, we have already shown that the trees do change over time. Figure \ref{July} shows the equivalent currency-tree one month later. The CAD cluster is still evident and, in fact, has strengthened: all nine CAD nodes are linked together. The NZD cluster is still evident, this time near to the top of the Figure. More interesting are the clusters which have changed. The CHF cluster has completely disintegrated; the CHF nodes are scattered over the tree. Hence the Swiss franc is no longer in play. Conversely, there is now a strong American dollar (USD) cluster which has formed, 
indicating that the dollar has become more important in determining currency moves. 

In short, it has become possibly to identify currencies which are actively in play and are effectively dominating the FX market. Sometimes, when currencies are in play, it will be obvious to traders: for example, when there is a large and sustained USD move. However, this is not always the case, and our currency-trees are able to provide an indication of how important (i.e. how much in play) a particular currency is. A crude measure of this could simply be the number of nodes which are connected together and which contain this currency. More sophisticated measures of the strength of the clustering, would involve looking at the underlying ultrametric hierarchy associated with the tree. Such measures will be investigated in future work, as will the temporal dynamics and `competition' between the various currency clusters within the tree. 

\section{VIII. Conclusions} 
We have provided a detailed analysis of the correlation networks, and in particular the Minimal Spanning Trees, associated with empirical data obtainined from the Foreign Exchange (FX) currency markets. This analysis has highlighted various data-related features which make this study quite distinct from earlier work on equities. 

We have shown that there is a clear difference between the currency trees formed from real markets and those formed from randomized data. Not only does the tree `look different', the degree distributions of the two trees are markedly different.  For the trees from real markets, there is a clear regional clustering. We have also investigated the time-dependence of the trees. Even though the market structure does change rapidly enough to identify changes in which currency-pairs are clustering together, there are links in the tree which last over the entire two year period. This shows that there is a certain robust structure to the FX markets.
We have also developed a methodology for interpreting the trees which has practical applications: the trees can be used to identify currencies which are in play. Whilst this does not have predictive power, it helps one to identify more accurately the state the market is currently in. Armed with this information, one can be more confident of the predictions made from other models. In future work, we will look at  trying to isolate the effect of news on the FX market -- in other words, the extent to which external news `shakes' the FX tree. Of particular interest is whether particular clusters have increased robustness over others, or not. In addition, we shall be investigating how tree structure depends on the frequency of the data used.


\begin{thebibliography}{99}
\bibitem{note9} We assume that the correlation structure is specified by the correlation coefficients.
\bibitem{note5} If there were no missing data points, this would give rise to a dataset of 4680 data points.
\bibitem{note6} This gave rise to the final dataset which contained 4608 datapoints. Therefore, the data cleaning procedure removed less than 2 percent of the 4680 possible datapoints.
\bibitem{note3} Even though quoting the two exchange-rates in this form seems slightly idiosyncratic, it is actually how the rates are quoted on the market. Most currencies are quoted with USD as the base currency. However GBP/USD is quoted in this unusual form because
historically GBP was not based on a decimal system.
\end{thebibliography}
\end{document}